\documentclass[twoside,leqno,twocolumn]{article}

\usepackage[letterpaper]{geometry}

\usepackage{graphicx}
\usepackage{ltexpprt}
\usepackage{amssymb}
\usepackage{amsmath}
\usepackage{xspace}
\usepackage{subcaption}
\usepackage{booktabs}
\usepackage{adjustbox}
\usepackage{floatrow}
\newtheorem{definition}{Definition}
\usepackage{stfloats}
\usepackage{hyperref}

\begin{document}

\newcommand\relatedversion{}
\renewcommand\relatedversion{\thanks{The full version of the paper can be accessed at \protect\url{https://arxiv.org/abs/1902.09310}}} 

\title{\textsc{FastCover}: An Unsupervised Learning Framework for Multi-Hop Influence Maximization
in 
Social Networks
}
\author{
Runbo Ni\thanks{
Department of Computer Science and Engineering, Shanghai Jiao Tong University, Shanghai, China.
\{pqross, karroyan, solour\_lfq\}@sjtu.edu.cn, \{gao-xf, gchen\}@cs.sjtu.edu.cn.
}
\and 
Xueyan Li\footnotemark[1]
\and
Fangqi Li\footnotemark[1]
\and
Xiaofeng Gao\footnotemark[1] \thanks{Corresponding Author.}
\and
Guihai Chen\footnotemark[1]
}

\date{}

\maketitle


\fancyfoot[R]{\scriptsize{Copyright \textcopyright\ 2022 by SIAM\\
Unauthorized reproduction of this article is prohibited}}





\newcommand{\model}{\textsc{FastCover}\xspace}
\newcommand{\kdDSP}{$k$-$d$DSP\xspace}
\newcommand{\mddsp}{M$d$DSP\xspace}

\begin{abstract}
Finding influential users in social networks is a fundamental problem with many possible useful applications. Viewing the social network as a graph, the influence of a set of users can be measured by the number of neighbors located within a given number of hops in the network, where each hop marks a step of influence diffusion.
In this paper, we reduce the problem of IM to a budget-constrained $d$-hop dominating set problem (\kdDSP). 
We propose a unified machine learning~(ML) framework, \textsc{FastCover}, to solve \kdDSP by learning an efficient greedy strategy in an unsupervised way. 
As one critical component of the framework, we devise a novel graph neural network (GNN) architecture, graph reversed attention network (GRAT), that captures the diffusion process among neighbors. 
Unlike most heuristic algorithms and concurrent ML frameworks for combinatorial optimization problems, \textsc{FastCover} determines the entire seed set from the nodes' scores computed with only one forward propagation of the GNN and has a time complexity quasi-linear in the graph size. 
Experiments on synthetic graphs and real-world social networks demonstrate that \textsc{FastCover} finds solutions with better or comparable quality rendered by the concurrent algorithms while achieving a speedup of over 1000x.
The source code of our model is available at \url{https://github.com/pqros/fastCover}.

\end{abstract}

\section{Introduction}
Identifying a group of participants that can influence as many other participants as possible, known as the influence maximization~(IM), is an extensively studied problem in social network analysis and has been widely applied to personalized recommendation~\cite{song2006personalized}, target advertisement~\cite{tang2018social}, and influential twitters recognition~\cite{mei2017influence}, etc. For instance, solving IM allows us to effeciently find potential brand spokespeople on social media to maximize the diffusion of information.




Specifically, a social network can be viewed as a graph with agents as nodes and inter-agent interactions as edges, along which the influence is diffused to peripheral agents.
Therefore, the problem of IM in social networks is essentially a dominating set problem (MDSP), in which we look for a subset of nodes whose influence covers the most agents through its neighborhood.
Traditional solutions to IM in social networks usually regard the one-hop coverage as deterministic or probabilistic diffusion models between direct neighbors such as \textit{independent cascade}~(IC) and \textit{linear threshold}~(LT) models.
In practice, we observe that interactions in social networks sometimes take the form of multiple hops as well.
For example, when recognizing influential twitters, not only the followers, but the followers' friends will also be influenced. 

\begin{figure}[ht]
    \centering
    \includegraphics[width=.9\textwidth]{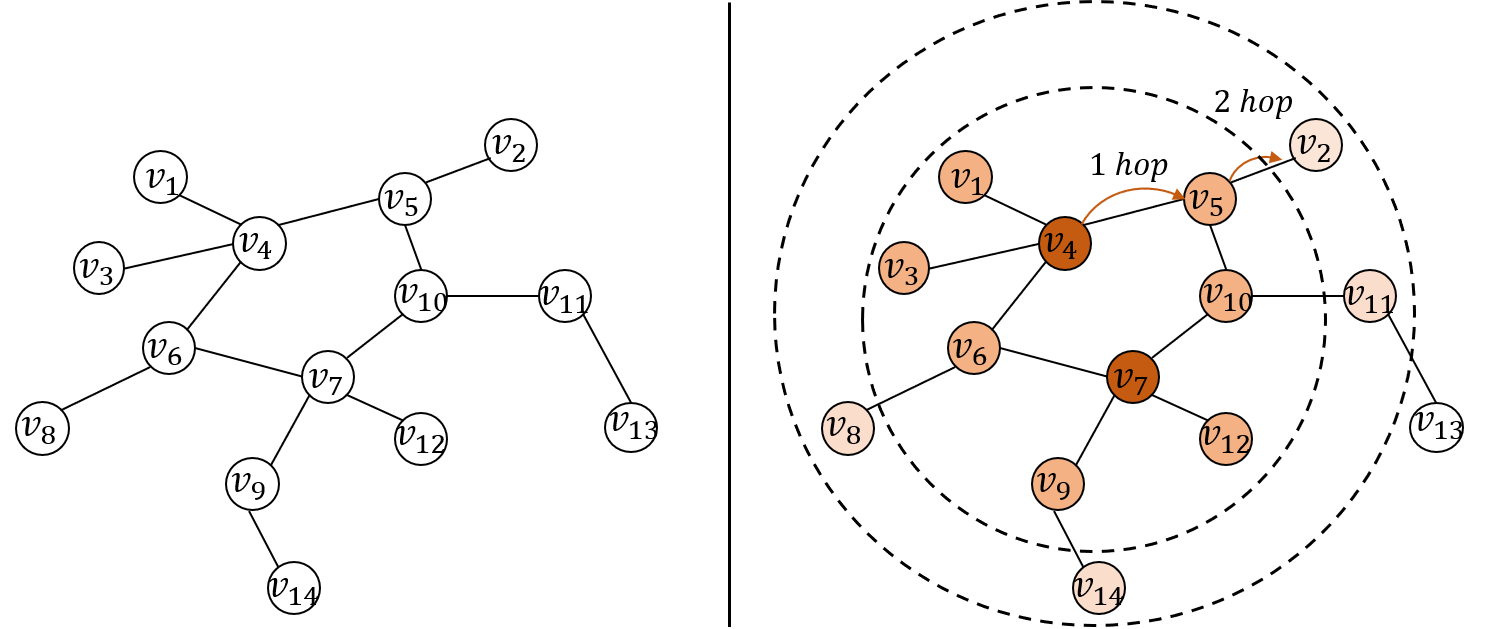}
    \caption{Illustration of multi-hop coverage in an undirected graph}
    \label{fig:toy_example}
\end{figure}

The $d$-hop dominating set problem (\mddsp) is a variant of MDSP that directly considers multi-hop coverage among nodes.
For a graph $G = (V, E)$, \mddsp aims at finding a $d$-dominating set $S$ in $G$, such that every node either belongs to $S$ or is reachable from nodes in $S$ within $d$ hops. 
In the toy example illustrated in Fig.~\ref{fig:toy_example} for example, $v_5$ is $1$-hop from $v_4$ and $v_2$ is $2$-hop from $v_4$. 
$S = \{v_4, v_7\}$ is not $2$-dominating since it takes 3 steps to cover $v_{13}$ from $S$. 

Exact solutions to \mddsp are expensive to obtain, especially for large-scale graphs.
The greedy algorithm is a benchmark heuristic for \mddsp, while computing multi-hop neighbors generates a prohibitive complexity. Studies on most \textit{machine learning} (ML)-based frameworks for combinatorial optimization are also limited to graphs containing fewer than tens of thousands of nodes.
With regard to real-world social networks, we need more efficient approaches to solve \mddsp with high quality.



In this paper, we present an unsupervised learning framework, \model, to solve $k$-budget-constrained \mddsp, namely \kdDSP. 
We propose, as part of the framework, a novel graph neural network (GNN) called \textit{graph reverse attention network} (GRAT) that incorporates the influence of the neighborhood into numerical features of each vertex.
\model is evaluated on synthetic graphs and real-world networks with up to 400k vertices. 
Empirical results prove that the model find comparable or even better solutions while achieving a significant speedup compared to state-of-the-art models.
Therefore, \model can serve as a competitive candidate for solving \kdDSP. 

The contribution of this work is threefold:

\begin{itemize}
    \item We propose a novel framework, \model, solving the budget-constrained \mddsp effectively and efficiently with attention-based graph neural networks in an unsupervised approach. 
    \item To better model the coverage, we devise graph reversed attention network (GART), a novel graph neural network architecture that leverages the inner characteristics of the diffusion process among neighbors.
    \item We conduct extensive experiments on synthetic graphs and real-world networks, demonstrating the superiority of our model in terms of the inference speed and solution quality.
\end{itemize}

The rest of this paper is organized in the following structure: Section \ref{sect:bg} introduces the background information and the related works. 
In Section \ref{sect:formulation}, we give the definitions and notations of the formulation of the $k$-budget-constraint version of the $d$-dominating set problem (\kdDSP). 
In Section \ref{sect:models}, we present the model \model that efficiently solves the problem.
In Section \ref{sect:exp}, we compare our model against concurrent solutions through experiments on synthetic and real-world datasets.
Finally, Section \ref{sect:conclusion} concludes the work.


\section{Related Works}
\label{sect:bg}
Among the studies on IM in social networks, many assume IC or LT as the underlying diffusion model. Goyal et al.~\cite{celf_www_2011} adopts the Monte Carlo simulation to estimate the influence ability of nodes and construct the seed set greedily. 
Tang et al. \cite{Tang} introduce martingales to study the random diffusion of propagation, and Nguyen et al. \cite{Nguyen} further accelerate their sampling algorithm. Chen et al.~\cite{DBLP:conf/sdm/ChenP19} propose a model which jointly models the effect of influence on both link and post generation.
However, both diffusion models neglect the direct diffusion over multi-hop neighbors, which is often observed in practical scenarios in service-oriented social networks.

IM with the multi-hop diffusion has been studied recently based on the dominating set problem, which aims at finding a subset of vertices covering the whole graph. 
Traditional algorithms such as \cite{10.1145/1557019.1557047} have guaranteed performance but are usually time-consuming
In particular, heuristic algorithms for \mddsp are commonly designed with greedy strategy. 
Basuchowdhuri and Majumder \cite{greedy-k-hop2014} propose a greedy heuristic algorithm for \mddsp that repetitively selects the node with maximum coverage among the remaining nodes. Nguyen et al. 
\cite{nguyen2020solving} pre-compute the multi-hop coverage of each node and add a pre-optimization and a post-optimization phase, which empirically improves the speed of \cite{greedy-k-hop2014}. 
However, such heuristic algorithms sacrifice the performance for time, motivating us to seek other methods to solve \mddsp. 



To the best of our knowledge, no ML-based solution has been proposed for \mddsp, while attempts are made to apply neural networks (NN) in solving other combinatorial optimization problems on graphs. 
A first demonstration is the application of Hopfield network \cite{wilson_stability_1988} in traveling salesman problem (TSP) by optimizing an energy function. 
Karalias and Loukas \cite{NEURIPS2020_49f85a9e} extends the idea of energy function to a more general form based on Erdõs probabilistic methods. 

As a popular alternative approach, \texttt{S2V-DQN} is a general RL-based framework for graph combinatorial optimization proposed by Khalil et al. \cite{NIPS2017_d9896106}. 
It uses GNN for graph embedding and \textit{deep Q-network} (DQN) for node selection strategy to form the solution and is widely adopted in later works on ML for combinatorial optimization. 
In particular, with regards to social influence, Mittal et al. \cite{NEURIPS2020_e7532dbe} incorporates several heuristics in the pre-processing phase for graph reduction in solving vertex cover and IM. 
Wang et al. \cite{DASFAA2021_wang} leverage user features in the graph embedding to solve IM in a social network. 
Though these RL-based solutions generate near-optimal solutions on various combinatorial optimization problems, one common bottleneck is the heavy time consumption of repetitive evaluations of neural networks, and they can hardly be generalized to larger graphs, such as social networks in reality.

\begin{figure*}[ht]
    \centering
    \includegraphics[width=.9\textwidth]{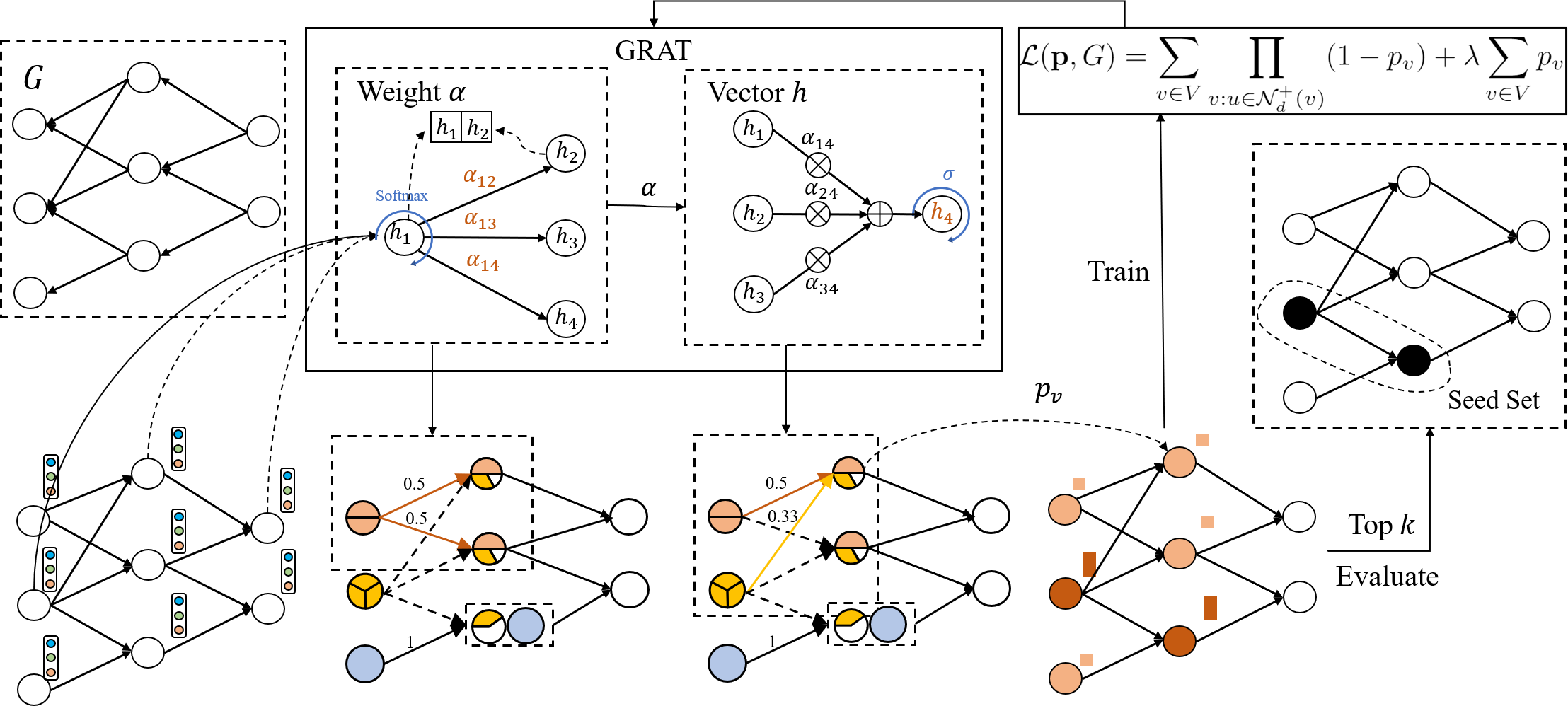}
    \caption{Illustration of the proposed framework \model as applied to a simple directed graph of eight nodes.}
    \label{fig:framework_v2}
\end{figure*}

\section{Problem Formulation} 
\label{sect:formulation}

\newcommand{\NN}{\mathcal{N}}
\newcommand{\NNp}{\mathcal{N}^{+}}
\newcommand{\dist}{\texttt{dist}}
In a directed/undirected graph $G = (V, E)$, the distance $\dist(u, v)$ from a vertex $u$ to $v$ is defined as the length of the shortest path from $u$ to $v$. When $U \subseteq V$, we define $\dist(U, v) \triangleq \min_{u \in U} \dist(u, v)$. 
\begin{definition}[$d$-coverage]
    The \textit{$d$-coverage} ($d \ge 0$)  of a vertex $u\in V$ denoted by $\NN^{+}_d (u)$ is the set of vertices reachable within $d$ steps from $u$, i.e.
\begin{equation*}
    \NN^{+}_d(u) \triangleq \left\{v \mid \dist(u, v) \le d \right\}.
\end{equation*} 
\end{definition}
If $v \in \NN^{+}_d(u)$ then we say that $v$ is \textit{$d$-covered} by $u$.
In particular, we have $\NN^{+}_0 (u) = \{ u \}$ and $\NN^{+}_1(u)$ coincides with $u$'s \textit{out-neighbors/successors} united with itself $\NN^{+}(u) \cup \{ u \}$. 
For $d > 1$, $\NN^+_d(u)$ can be recursively computed with
\begin{equation*}
    \NN^+_d(u) = \NN_{d-1}^{+}(u) \cup \left(\bigcup_{ v\in \NN_{d-1}^+(u)}\NN_1^+(v)\right).
\end{equation*}
Likewise, we define the \textit{$d$-coverage} of a set $U \subseteq V$ by $\NN^{+}_d(U) = \bigcup_{u \in U} \NN^{+}_d(u)$. 
A $d$-dominating set of $G$ is a subset $S \subseteq V$ whose $d$-coverage is $V$, and \cite{greedy-k-hop2014} defines the minimum $d$-dominating set problem (\mddsp) as to find a $d$-dominating set with the minimum cardinality. 
Here we define the $k$-budget-constrained version of the $d$-dominating set problem (\kdDSP) as follows. 

\begin{definition}[\kdDSP]
    Given $G = (V, E)$, the budget $k$ and the maximum hop count $d$, the goal is to find a subset $S \subseteq V$ with cardinality at most $k$ (i.e., $|S| \le k$) such that the number of nodes $d$-covered by $S$ is maximized.
\end{definition}
Here, the subset $S$ is also referred to as the \textit{seed set} and nodes in $S$ are \textit{seeds}. Alternatively, \kdDSP can be formally defined as the optimization problem \eqref{eq:kdDSP}:
%
%
\begin{equation}
    \begin{aligned}
        &\text { max }_{S \subseteq V} &\left|\mathcal{N}_d^+(S)\right| \\
        &\text { subject to } &|S| \le k.
    \end{aligned}
    \label{eq:kdDSP}
\end{equation}

As pointed out in \cite{greedy-k-hop2014}, \mddsp is $\mathcal{NP}$-hard for all $d \ge 1$.
Since \kdDSP contains \mddsp as a particular case, its $\mathcal{NP}$-hardness holds as well.
%
%


\section{Proposed Model}
\label{sect:models}

In this section, we present our framework \model that solving \kdDSP, 
which is graphically illustrated in Fig. \ref{fig:framework_v2}.

\subsection{Motivation}
\label{sect: motivation}

The greedy heuristic is widely applied in solving graph optimization as well as coverage problems.
\cite{greedy-k-hop2014} proposes a simple but competitive greedy algorithm \texttt{Greedy} solving \mddsp based on the \textit{effective coverage} of each vertex $u \in V$, i.e., the number of nodes in $\NN_d^+\left({u}\right)$ not $d$-covered by already selected seeds.

Despite its privileges in the quality of solutions, \texttt{Greedy} has prohibitive running time on large graphs for $d > 1$ due to two bottlenecks:
(1) computing the $d$-coverage of each node; 
(2) updating the effective coverage of each node after updating the seed set. 

\newcommand{\rrrr}[1]{\textcolor{red}{[TODO: #1]}}


To overcome these limitations, we want to assign a \textit{score} to each node different from the effective coverage, representing its ``overall'' influence to embed the topology of the graph. In particular, we hope that such score enables us to obtain high-quality solutions by simply picking the nodes with top-$k$ scores as the seed set with budget $k$, and we can ideally circumvent the second bottleneck.
Moreover, if the scores of all nodes can be calculated in linear time with respect to the graph then the total time complexity can be effectively reduced to $\mathcal{O} \left( |E| + |V| \log |V| \right)$.

However, it is not easy to define \textit{score} explicitly. One plausible definition of the score for a node $u$ is the cardinality of its $d$-coverage $\left| \NN_d^{+} (u) \right|$. 
However, such heuristic is flawed due to the \textit{similar point effect}. 
Two nodes are similar if they have a highly overlapping $d$-coverage, and including both of them in the seed set can be a waste of budget. 
Such similarity of two nodes $u$ and $v$ can be measured with the Jaccard similarity of their $d$-coverage defined as
$\mathcal{J}_d(u, v) \triangleq \frac{|\NN^{+}_d(u) \bigcap \NN^{+}_d(v)|}{|\NN^{+}_d(u) \bigcup \NN^{+}_d(v)|}.$

\begin{figure}[ht] 
\centering
\begin{subfigure}[t]{.4\columnwidth}
\centering\includegraphics[width=.9\linewidth]{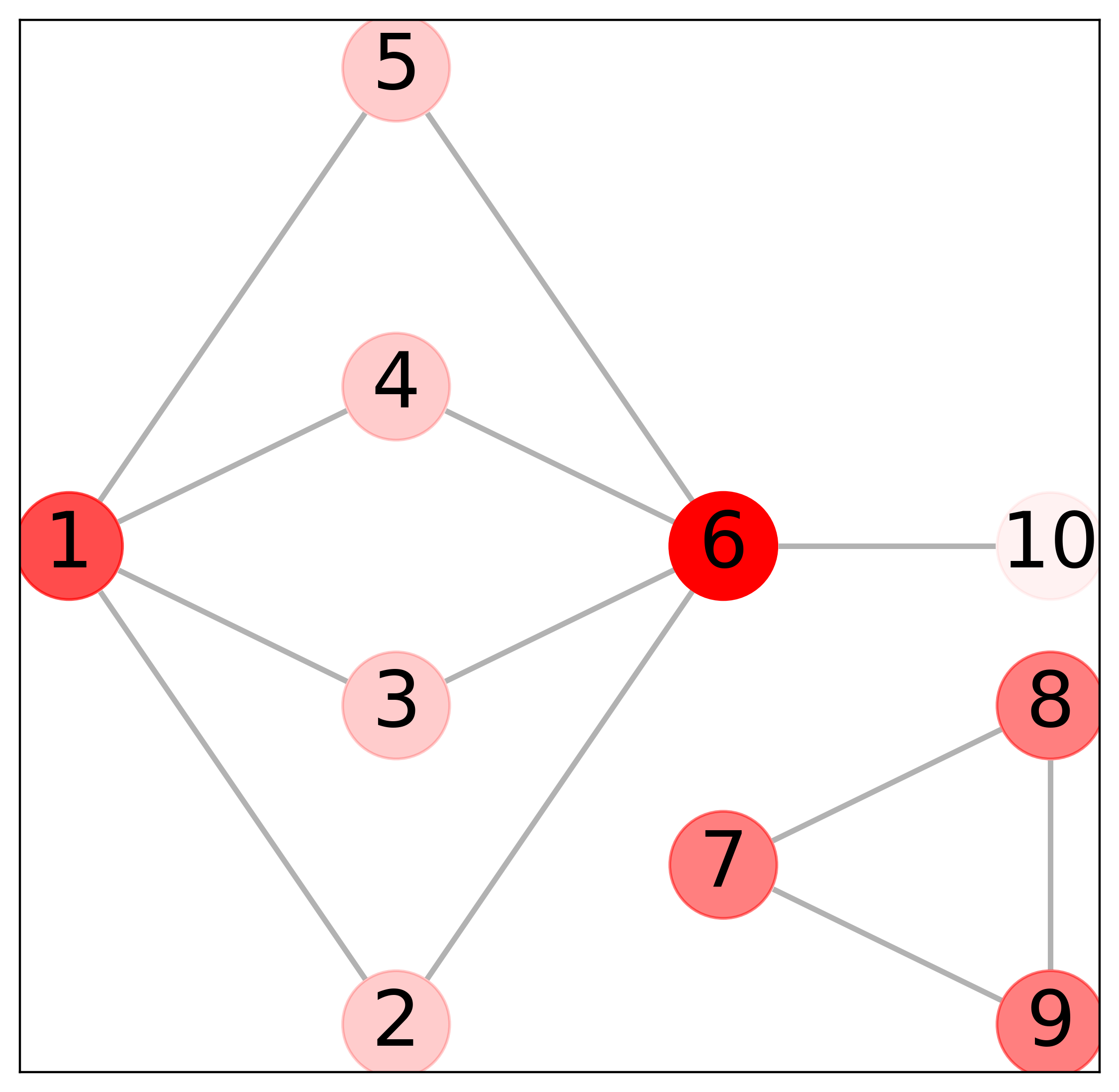}
\caption{Toy example}
\label{fig:symm_toy}
\end{subfigure}
\begin{subfigure}[t]{.5\columnwidth}
\centering\includegraphics[width=\linewidth]{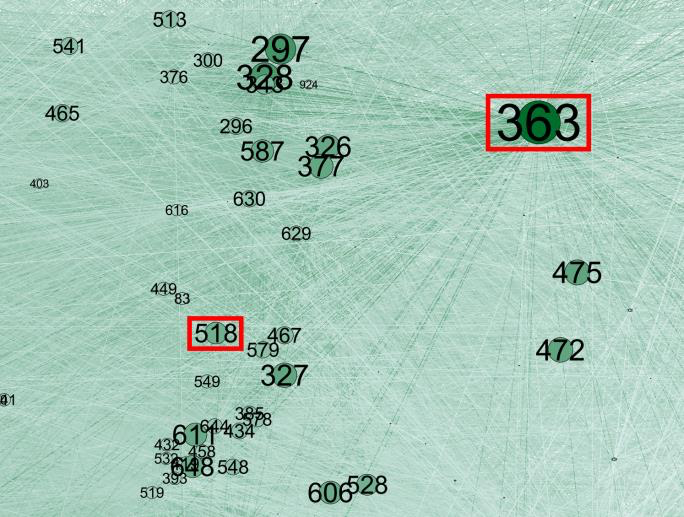}
\caption{HepPh}
\label{fig:symm_hep}
\end{subfigure}
\caption{Illustration of similar point effect in toy example and real graph.}
\label{fig:symm}
\end{figure}

Consider the toy example in Fig. \ref{fig:symm_toy} with $d = 1$ and $k = 2$. 
Node 6 and 1 are the nodes with the most neighbors, but they have a significant Jaccard similarity of $0.83$. Given the node 6 already selected as a seed, the optimal choice of the second seed should be 7, 8, or 9 instead of 1.
Similar phenomena are shared in real graphs. 
In HepPh~\cite{hepph} illustrated in Fig.~\ref{fig:symm_hep}, node 328 and 297 have respectively 486 and 482 neighbors, ranked the second and third among all the vertices. 
They have 450 neighbors in common with $\mathcal{J}_1(328, 297) = 86.9\%$. 
As a result, we have to find a score that better reflects the relation among nodes close to each other, such as the overlapping effect between different nodes' coverage.

In particular, GNN is a candidate solution to learn an ideal score implicitly.

\subsection{Top-level Framework Design}

Based on the analysis in Sect.~\ref{sect: motivation}, we propose \model, an unsupervised learning model incorporating a GNN that computes the score of nodes and solve \kdDSP efficiently.
In general, a GNN captures the local interaction between neighbor nodes in a directed/undirected graph and finds a representation for each node. 
Typical GNN structures such as graph convolutional network (GCN)~\cite{kipf_semi-supervised_2017} and graph attention network (GAT)~\cite{velickovic_graph_2018} have a time complexity of $\mathcal{O} \left(C (|E| + |V|) \right)$ for one forward propagation, where $C$ is determined by the GNN and is independent of the input graph.
Therefore, GNN can efficiently incorporate the neighborhood information and evaluate nodes.

The major components of \model illustrated in Fig.~\ref{fig:framework_v2} are described as follows:


\begin{itemize}
    \item \textit{Pre-processing}. The direction of all the edges in the graph is \textit{reversed}, and the vertices are embedded as numerical vectors and fed into the GNN.
    \item \textit{Score}. A multi-layer GNN called \textit{graph reversed attention network} (GRAT) maps each node to its score bounded $[0, 1]$.
    \item \textit{Loss Function}. In the training phase, parameters in the GNN are optimized through a differentiable loss function over the nodes' scores. 
\end{itemize}
 
In the evaluation phase, we simply compute each node's score by the trained GNN with one forward pass and construct the seed set with a top-$k$ search.

\subsection{Differentiable Loss Function}

\newcommand{\pp}{\mathbf{p}}
Inspired by randomized algorithms, we interpret the scores in a probabilistic approach. 
We want the scores $\pp = (p_v)_{v\in V}$ to represent the probability that each node $v$ is selected as part of the seed set. As we do not \textit{a priori} impose the budget $k$, we only focus on the relative amplitude of the probabilities among the nodes.
The loss function on the scores $\pp$ of the nodes in the graph $G$ is defined as follows:
\begin{equation}
    \mathcal{L}(\pp, G) \triangleq \mathbb{E} \left[|\text{uncovered vertices}|\right] + \lambda \mathbb{E} \left[|\text{seed set}|\right],
    \label{eq:loss_raw}
\end{equation}
where $\lambda > 0$, and the expectations are computed by integrating out the randomness in selecting the seed set w.r.t. $\pp$.
The first term in \eqref{eq:loss_raw} penalizes the expected number of uninfluenced vertices. 
Minimizing this term is tantamount to maximizing the number of $d$-covered vertices. 
The second term regularizes the seed set's expected size so that vertices with less importance are less likely to be picked. 
Without the latter term, the optimal solution of \eqref{eq:loss_raw} is obviously $p_v = 1$ for all $v\in V$. 
$\lambda$ balances the impact of these two terms.

In order to express \eqref{eq:loss_raw} explicitly, we further suppose that the nodes are included in the seed set independently.
By the linearity of the expectation, the expected cardinality of the seed set is $\sum_{v\in V}p_v$. 
A vertex $u$ being not covered is equivalent to the event that neither the vertex itself nor any vertex having $u$ in its $d$-coverage being included in the seed set, i.e.
\begin{equation*}
    \mathbb{P}\left(u~\text{is~not~covered} \mid \pp, G\right) = \prod_{v : u \in \mathcal{N}^{+}_d(v)} (1 - p_v), \quad u\in V.
\end{equation*}
Hence the loss function can be explicitly written as
\begin{equation}
    \mathcal{L}(\pp, G) = \sum_{v\in V} \prod_{v : u \in \mathcal{N}^{+}_d(v)} (1 - p_v) + \lambda \sum_{v\in V}p_v.
    \label{eq:loss}
\end{equation}
Notice that
\begin{itemize}
    \item $\mathcal{L}(\cdot, \cdot)$ is only evaluated in the training phase of the GNN. So its space complexity does not hinder the efficiency of the inferring phase.
    \item The product $\prod_{v : u \in \NN_d^+(v)} (1 - p_v)$ in \eqref{eq:loss} can be fully vectorized with the adjacency matrix of the reversed graph $G^{rev}$. Thus, the evaluation of $\mathcal{L}(\cdot, \cdot)$ and its gradients can be paralleled and accelerated with GPU.
\end{itemize}

Additionally, we observe that the problem of similar point effect is effectively alleviated with \eqref{eq:loss}. 
With $d = 1$, the node 328 and 297 of HepPh, ranked $2$nd and $3$rd in terms of their out-degrees, have scores respectively ranked $2$nd and $494$th by minimizing \eqref{eq:loss} with $\lambda = 1$.

\subsection{GNN Architecture}

Directly optimizing \eqref{eq:loss} with respect to $\pp$ is very time-consuming, especially on large graphs. 
So we intend to use a GNN to find an approximate solution.


\subsubsection{Input}
\newcommand{\rev}{\text{rev}}
In \kdDSP, the input is a graph $G = (V, E)$. 
We first transform $G$ into its reversed graph $G^{\rev}$ by flipping all the edges' direction.
The intuition is that a node's score/importance is determined by its successors in $G$ in our context, and will be further explained in the GRAT part.
Uniform initialization is adopted for the node embedding in our model since we only leverage the network topology.



\subsubsection{GRAT Layer}
To incorporate the characteristics of \kdDSP, we design a novel GNN architecture -- \textit{graph reversed attention network} (GRAT), in which we integrate the attention mechanism~\cite{velickovic_graph_2018} 
at the source nodes instead of the destination nodes in traditional GAT.

As typical GNNs, the input of a GRAT layer is a set of node features $\left(h^{(l)}_{v} \right)_{v\in V}$ and the output is $\left(h^{(l+1)}_{v} \right)_{v\in V}$. 
In the message-passing process, an edge from $u$ to $v$ has its feature expressed as the concatenation of its incident vertices $u$ and $v$
\begin{equation*}
    e_{uv}^{(l)} = [h_u^{(l)} | h_v^{(l)}].
\end{equation*}

For $u\in V$, the attention coefficients are computed over $\mathcal{N}^{+}(u)$ 
by
\begin{equation}
    \alpha_{uv}^{(l)} = \dfrac
    {\exp\left( \operatorname{ReLU} \left( a^{(l)^{\top}} e_{uv}^{(l)} \right) \right)}
    {\sum\limits_{w \in \NN^{+}(u)} \exp\left( \operatorname{ReLU} \left( a^{(l)^{\top}} e_{uw}^{(l)} \right) \right)}, \quad v\in \mathcal{N}^{+}(u),
    \label{eq:attention}
\end{equation}
where the activation function $\operatorname{ReLU}(\cdot) = \max (\cdot, 0)$, and $a^{(l)}$ is a trainable vector.

Finally, the output of the GRAT layer is given by
\begin{equation}
    h^{(l+1)}_{u} = \sigma \left(  \sum_{w \in \NN^{-}(u)} \alpha_{wu}^{(l)} W^{(l)} h^{(l)}_{w} + b^{(l)}\right),\quad u\in V,
    \label{eq:attention_next}
\end{equation}
where $\NN^{-}(u)$ denotes the predecessors of $u$, and $W^{(l)}$, $b^{(l)}$ are trainable parameters. $\sigma(\cdot)$ is an optional activation function.

\begin{figure}[ht]
    \centering
    \begin{subfigure}[t]{.45\columnwidth}
    \centering\includegraphics[height=3cm]{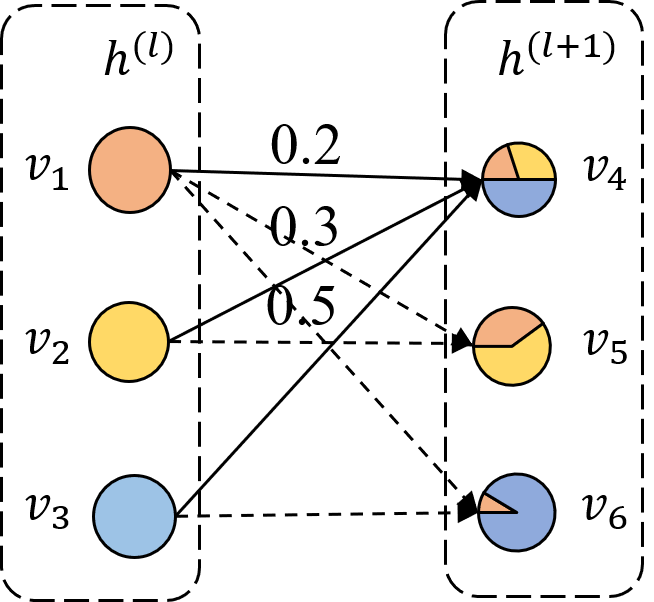}
    \caption{GAT}
    \label{fig:gat}
    \end{subfigure}
    \begin{subfigure}[t]{.45\columnwidth}
    \centering\includegraphics[height=3cm]{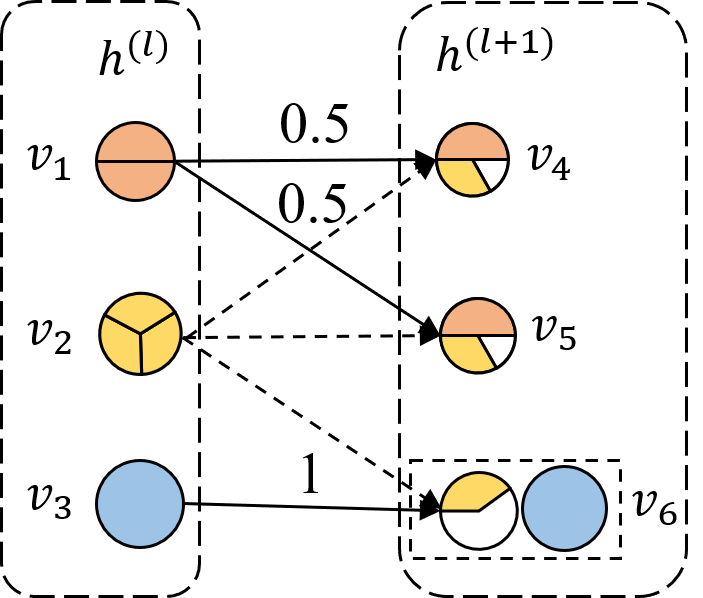}
    \caption{GRAT}
    \label{fig:grat}
    \end{subfigure}
    \caption{Illustration of updating $h^{(l+1)}$ for $v_{4}, v_{5}, v_{6}$ from $h^{(l)}$ on $v_{1},v_{2},v_{3}$ using GRAT and GAT. Self-loops are omitted.
    In (a), the attention coefficients of a node's sources are normalized in GAT. 
    In (b), the attention coefficients in GRAT partition each source node's feature over its successors, and the total weight received at a destination node can be other than 1. \\
    }
    \label{fig:grat_vs_gat}
\end{figure}



Fig. \ref{fig:grat_vs_gat} illustrates differences between GRAT and the classic GAT. 
For a node $u$, 
The attention coefficients in GAT are normalized over its predecessors $\NN^{-}(u)$ so that the updated features are comparable across different nodes \cite{velickovic_graph_2018}.
In \kdDSP, our goal is to identify the difference in the influence among the nodes easily, and such a normalization mechanism in GAT may not be the most appropriate.
In GRAT, instead, the attention coefficients are normalized for each source node $u \in V$ over its successors $\NN^{+}(u)$. 

The motivation behind GRAT is as follows. Consider the simplest case with $d = 1$. 
A node being $d$-covered by multiple nodes does not increase the reward. 
Therefore we interpret $h^{(l)}(u)$ as the reward signal of covering $u$ and distribute it through attention coefficients in GRAT across $\NN^{+}(u)$, i.e. the vertices covering $u$ in $G$ as GRAT is applied to $G^{\text{rev}}$. 
As a result, the nodes with more extensive coverage are likely to receive a stronger reward signal and potentially a better score. 
Also, if a node's coverage overlaps with other nodes, the total reward it receives should decline due to the attention mechanism, which may alleviate similar point effects.

In particular, with three hidden layers of 32 hidden units each, \model using GRAT assigns significantly different scores to the similar points 328 and 297 in HepPh, while that using GAT cannot effectively distinguish them. 


\subsection{Complexity Analysis}
To analyze the complexity of the proposed model, we first compute the number of parameters of GRAT in \model. 
A GRAT layer with $n_{i}$ input features and $n_{o}$ output features contains $\left( n_i n_o + 2 n_i + n_o \right)$ trainable parameters in $a$, $W$ and $b$. Therefore, a $r$-layer GRAT network contains
$\sum_{i = 1}^{r} \left( n_i n_{i+1} + 2 n_i + n_{i+1} \right)$
trainable parameters, where the $i^{\text{th}}$ layer ($i \in [1,\cdots,r]$) contains $n_{i}$ hidden units, and $n_{r+1} = 1$ is the scalar output.

As for the time complexity, computing \eqref{eq:attention} and \eqref{eq:attention_next} in a GRAT layer has respectively complexity of $\mathcal{O}\left( n_i |E| \right)$ and $\mathcal{O}\left( n_i n_o |V| + n_o |E| \right)$. 
Thus, the total time complexity evaluating GRAT is linear in the graph size, which is on par with GAT and GCN. 
Finally, we conduct top-$k$ sort on the output, so the total complexity is $\mathcal{O}\left( |V| \log |V| + |E| \right)$, lower than \texttt{Greedy}. 

In particular, when $n_i$'s are constantly $n$, the number of parameters is $ r n^2 + (3r - 1) n - 1$ and the time complexity for each layer is of order $\mathcal{O}(n^{2}|V|+n|E|)$.



\section{Experimental Evaluation}
\label{sect:exp}

In this section, we first look at the performance of candidate GNN architectures in \model. 
Then we compare the optimally configured \model against concurrent heuristic or ML-based algorithms in solving \kdDSP through experiments on synthetic and real-world networks. 
We empirically show that our model provides a comparable or better solution to the problem while being significantly faster. 
The source code is found at \url{https://github.com/pqros/fastCover}.

\label{sect:experiments}

\subsection{Experiment Setup}

\subsubsection{General Settings}
All experiments are conducted on a machine running Intel(R) Core(TM) i3-10100F CPU @ 3.60GHz with eight cores with a Nvidia 2080 Ti GPU. 
The results are obtained with five repeated experiments, and the sample mean of the metrics is reported.
Graph networks and algorithms are implemented in Python with libraries \texttt{igraph} and \texttt{dgl} with \texttt{PyTorch} backend, which matches the environment in the related works \cite{greedy-k-hop2014,NIPS2017_d9896106,nguyen2020solving}.


\subsubsection{Data}

We use both synthetic and real-world datasets to evaluate our work. 
The training and validation are carried on random Erdös-Renyi (ER)~\cite{erdHos1960evolution} graphs and tests are on both synthetic graphs and real social networks containing up to $4.0 \times 10^5$ nodes.

\subsection{Baseline Methods}
\label{sect:baselines}

The concurrent methods against which \model is compared in solving the problem of \kdDSP are described as follows and the time complexity of all the algorithms is summarized in Table \ref{tab:complexity}, where $n_d$ denotes the average cardinality of all nodes' $d$-coverage. 
\begin{itemize}
    \item \texttt{Greedy}: In each iteration, \texttt{Greedy} \cite{greedy-k-hop2014} adds the node with the maximum effective $d$-coverage to the seed set, and updates the effective coverage with \textit{breadth-first search} (BFS) .
    \item \texttt{CELF}: We modify \texttt{CELF}\cite{celf_www_2011} as accelerated implementation of \texttt{Greedy} for the problem of \kdDSP in two aspects. First, we pre-compute and store the $d$-coverage of all the nodes for $d > 1$ to avoid running BFS repeatedly. Second, we use a priority queue to store the effective $d$-coverage and applies ``lazy forward'' mechanism in the original \texttt{CELF} to avoid unnecessary updates. As a result, \texttt{CELF} largely accelarates \texttt{Greedy}~\cite{greedy-k-hop2014}.
    \item \texttt{Greedy-1}: Regardless of the value of $d$, \texttt{Greedy-1} iteratively selects the node with maximum effective $1$-coverage. We also apply the priority and lazy forward in \texttt{CELF}.
    \item \texttt{HEU}: \texttt{HEU} \cite{nguyen2020solving} is a light-weight 3-phase algorithm which integrates multiple heuristics based on the observed features of real graphs. 
    \item \texttt{S2V-DQN}: \texttt{S2V-DQN}~\cite{NIPS2017_d9896106} is a framework solving graph combinatorial problems with reinforcement learning, which uses a multi-layer GNN for graph and node embedding and learns a greedy strategy for the iterative node selection based on DQN.
\end{itemize}
\begin{table}[ht]
  \centering
  {
  \setlength{\tabcolsep}{0.65em}
  \begin{adjustbox}{width=.9\textwidth}
  \begin{tabular}{lcc}
    \toprule
    \textbf{Model} & \textbf{Complexity ($d = 1$)} & \textbf{Complexity ($d \ge 2$)} \\
    \midrule
    \texttt{Greedy} \cite{greedy-k-hop2014} & $\mathcal{O}(k|V| + |E|)$ & $\mathcal{O}(k |V| n_d)$  \\
    \texttt{CELF}            \cite{celf_www_2011}  & $\mathcal{O}(|E| \log |V|)$ & $\mathcal{O}(|V| n_d)$  \\
    \texttt{Greedy-1} \cite{greedy-k-hop2014} & $\mathcal{O}(|E| \log |V|)$ & $\mathcal{O}(|E| \log |V|)$ \\
    \texttt{HEU}             \cite{nguyen2020solving} & $\mathcal{O}\left(|E| + |V| \log |V|\right)$ & $\mathcal{O}\left(k |E| + |V| \log |V| \right)$   \\
    \texttt{S2V-DQN}         \cite{NIPS2017_d9896106} & $\mathcal{O}(k|E|)$ & $\mathcal{O}(k|E|)$ \\
    \textbf{\texttt{FastCover}} & $\mathcal{O}(|E| + |V| \log |V|)$ & $\mathcal{O}(|E| + |V| \log |V|)$ \\
    \bottomrule
  \end{tabular}
  \end{adjustbox}
  }

  \caption{Summary of time complexity of algorithms solving \kdDSP}
  \label{tab:complexity}
\end{table}

Among these algorithms, \model and \texttt{S2V-DQN} use GNN for node embedding, and we neglect the constant factor determined solely by their GNN architecture. 
The worst-case complexity of \model, \texttt{S2V-DQN} and \texttt{Greedy-1} is invariant in $d$, since these methods do not  nodes' $d$-coverage explicitly.
On the contrary, \texttt{Greedy}, \texttt{HEU} and \texttt{CELF} involve evaluations of the nodes' $d$-coverage, which significantly increase the complexity for $d > 1$.


As for the quality of solution, \texttt{CELF} (or \texttt{Greedy}) guarantees an approximation ratio of $\left(1 - 1/e \right)$ \cite{im_kdd_03}, while the other methods provide no theoretical guarantee. In practice, \texttt{CELF} even provides an empirical approximation ratio of over $99\%$ in solving $k$-maximum vertex cover problem \cite{NEURIPS2020_e7532dbe}. Therefore, the solution found by \texttt{CELF} can be approximately viewed as optimal.

Finally, for the memory consumption, \texttt{CELF} requires $\mathcal{O} (|V| n_d)$ space to store the $d$-coverage of all nodes for $d > 1$, while the other methods have all a space complexity linear in the graph size. In practice, $|V| n_d$ is usually significantly larger than $|E|$ due to the sparsity of real social networks. 

\subsection{Implementation Details of \model}
 
In \model, the GNN consists of $3$ layers of 32 hidden units per layer, with \texttt{ReLU} as the activation function for all but the last one, in which \texttt{sigmoid} is applied to normalize the results.
The training set consists of 20 directed ER graphs $n = 1000, p = 10/n$ with 15 training instances and 5 validation instances.
Each model is trained for up to 20 epochs using early stopping with patience 5 based on the average coverage rate in the validation set with a fixed $k$ for each $d = 1, 2, 3$. We set the maximal hop number as 3 since (1) it matches concurrent works \cite{greedy-k-hop2014,nguyen2020solving}; (2) the diameter of real social networks is usually limited.
The hyper-parameter $\lambda$ in the loss \eqref{eq:loss} is simply set as $1$. 


\begin{figure*}[htbp]
    \centering
    \includegraphics[width=.85\textwidth]{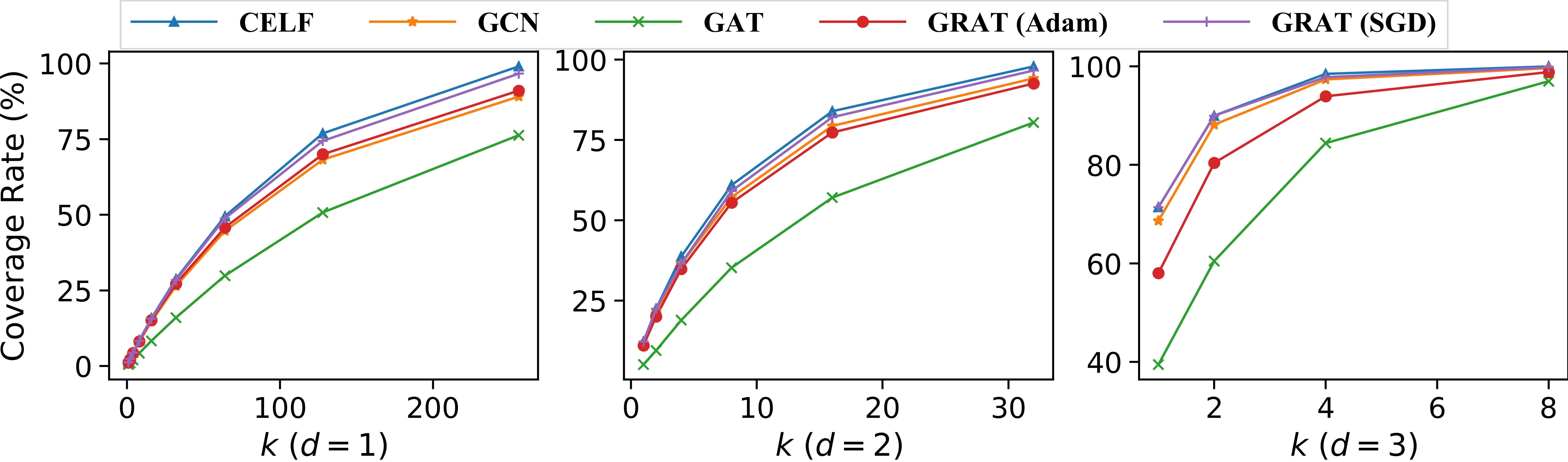}
    \caption{Coverage rate of \textsc{FastCover} using GRAT, GCN and GAT on ER-1000}
    \label{fig:perf1000}
\end{figure*}

\subsection{Results}
\subsubsection{Impact of GNN Architecture in \model}

We first study the influence of GNN architecture in \model by comparing the quality of solutions. For all the models, the training time is within 300 seconds, with 6-8 epochs to trigger the early stopping. In Fig. \ref{fig:perf1000}, we show the average coverage rate on 10 ER-1000 graphs held out in the training phase, and all GNNs are trained with stochastic gradient descent (SGD) unless otherwise annotated. We find that GRAT consistently outperforms GCN and GAT in terms of the coverage rate, and achieves a performance comparable with \texttt{CELF}. 

It is also interesting to find that \texttt{Adam} produces a lower coverage rate of \model with GRAT, which is potentially due to the better generalization capacity of the optimizer \texttt{SGD} compared to \texttt{Adam} \cite{NEURIPS2020_f3f27a32}. 

\model with GCN has an average coverage rate slightly below GRAT for $d = 2, 3$, while it renders significantly better performance than GAT, which suggests that the attention mechanism in GAT is limited in \kdDSP. On average, with $k$ ranging from $1$ to $128$, the coverage rate of GRAT is $99.0\%$ of \texttt{CELF} compared to $92.3\%$ and $62.9\%$ for GCN and GAT respectively .

We also notice that the performance of \model with GRAT is very close to \texttt{CELF} when the budget $k$ is limited. For each instance, GRAT makes the same choice as CELF on the first four candidates, while the following can largely differ. One possible explanation is that the loss function \eqref{eq:loss} is more sensitive to the score of nodes with strong influence, while the difference between nodes with less importance is not always correctly recognized. 

\begin{table}[ht]
    \centering
    \begin{adjustbox}{width=\textwidth}
    \setlength{\tabcolsep}{0.3em}
    \begin{tabular}{lrrrr|rrrr|rrrr}
    \toprule
    {} & \multicolumn{4}{c}{($d = 1, k = 64$)} & \multicolumn{4}{|c|}{($d = 2, k = 16$)} & \multicolumn{4}{c}{($d = 3, k = 4$)} \\
    \textbf{Model} 
    & \texttt{GCN} & \texttt{GAT} & \texttt{CELF} & \texttt{GRAT} 
    & \texttt{GCN} & \texttt{GAT} & \texttt{CELF} & \texttt{GRAT} 
    & \texttt{GCN} & \texttt{GAT} & \texttt{CELF} & \texttt{GRAT}           \\
    \midrule
    \textbf{ER-1k} 
    &  0.68             & 0.51      & \textbf{0.77}     & 0.75 
    &  0.94             & 0.80      & \textbf{0.98}     & 0.97 
    &  \textbf{1.00}    & 0.97      & \textbf{1.00}     & \textbf{1.00}     \\
    \textbf{ER-2k} 
    &  0.45             & 0.30      &   \textbf{0.49}   & \textbf{0.49} 
    &  0.79             & 0.57      &   \textbf{0.84}   & 0.82
    &  0.97             & 0.84      &   \textbf{0.98}   & \textbf{0.98}     \\
    \textbf{ER-4k} 
    &  0.26             & 0.16      &   \textbf{0.29}   & \textbf{0.29} 
    &  0.57             & 0.34      &   \textbf{0.61}   & 0.60
    &  0.87             & 0.63      &   \textbf{0.90}   & 0.88              \\
    \textbf{ER-8k} 
    &  0.15             & 0.08      &   \textbf{0.16}   & \textbf{0.16} 
    &  0.36             & 0.19      &   \textbf{0.39}   & \textbf{0.39} 
    &  0.68             & 0.41      &   \textbf{0.72}   & 0.70              \\
    \bottomrule
    \end{tabular}
    \end{adjustbox}
    \caption{Average coverage rate of \textsc{FastCover} using GRAT, GCN and GAT on ER-1000, ER-2000, ER-4000, ER-8000 in the test set.}
    \label{tab:syn_k}
\end{table}

We also randomly generate 10 ER graphs with $n=2000$, $4000$ and $8000$ with $p = 10/n$ to examine the generalization performance of \model. We fix $k = 64, 16, 4$ respectively for $d = 1, 2, 3$ and report the average coverage rates in Table \ref{tab:syn_k}. In all of the instances, \model with GRAT consistently outperforms GCN and GAT, always with gap within $2\%$ compared to CELF, which proves the effectiveness of \model and GRAT in solving \kdDSP.


\subsubsection{Real Graphs}
We fix the GRAT parameters in \model trained on ER graphs and benchmark its performance against concurrent methods on $6$ real social network graphs, whose number of vertices ranges from $5.9 \times 10^3$ to $4.0 \times 10^5$.

In Table \ref{tab:real_k_64}, we report the coverage rate on real networks with budget $k$ fixed to be $64$, as well as the number of vertices $n$ and edges $m$. Our method \model solves all the instances for d = 1, 2, 3 within $t=900s$ and achieves the highest rate in all but one case. The benchmark \texttt{CELF} has a stable performance when the algorithm terminates, while it fails in 5 cases for $d = 2, 3$. 

As the other GNN-based method, \texttt{S2V-DQN} cannot scale to the two largest graphs with over $10^5$ nodes due to the time limit of 900s. \texttt{GREEDY-1} and \texttt{HEU} are both light-weight heuristics that solve all the instances of \kdDSP. However, their performance can be largely inferior to \texttt{CELF} in some cases. 
Accordingly, we conclude that GRAT well generalizes to larger real graphs based on its coverage comparable to \texttt{CELF}.

\begin{table*}[t]
    \centering
    \scalebox{0.8}{
    \begin{tabular}{lcc|cccc|ccccc|ccccc}
    \toprule
    \multicolumn{3}{c}{} & \multicolumn{4}{c}{($d=1$)} & \multicolumn{5}{|c|}{($d=2$)} & \multicolumn{5}{c}{($d=3$)} \\
    \textbf{Graph}              &  
    $n$ & $m$ &
    \texttt{HEU}        & \texttt{DQN}      &  \texttt{CELF}    &  \texttt{FC}    &  
    \texttt{GR-1}       &  \texttt{HEU}     &  \texttt{DQN}     &  \texttt{CELF}    &  \texttt{FC}    &
    \texttt{GR-1}       &  \texttt{HEU}     &  \texttt{DQN}     &  \texttt{CELF}    &  \texttt{FC}    
    \\
    \midrule
    \textbf{sign-bitcoinotc} &  
    $5.9e3$ & $3.6e4$ &
    0.42                &    0.68           &   \textbf{0.69}   & \textbf{0.69}     & 
    0.97                & 0.85              &    0.97           & \textbf{0.98}     &   \textbf{0.98}   &     
    0.99                & 0.98              &    0.99           &   \textbf{1.00}   & \textbf{1.00} 
    \\
    \textbf{anybeat}         & 
    $1.3e4$ & $6.7e4$ &
    0.47                &    0.71           &   \textbf{0.72}   & \textbf{0.72}     &
    0.89                &   0.84            &    0.89           &   \textbf{0.90}   & \textbf{0.90} &     
    0.91                & 0.91              &    0.91           &   \textbf{0.92}   & \textbf{0.92} 
    \\
    \textbf{advogato}        & 
    $6.6e3$ & $5.1e4$ &
    0.28                &    0.43           &   \textbf{0.45}   & \textbf{0.45}     &
    0.63                & 0.60              &    0.64           &   \textbf{0.65}   & \textbf{0.65} &     
    0.65                & 0.66              &    0.66           &   \textbf{0.67}   & \textbf{0.67} 
    \\
    \textbf{slashdot}        & 
    $8.2e4$ & $5.5e5$ &
    \textbf{0.07}       &    \textbf{0.07}  &   \textbf{0.07}   & \textbf{0.07}     &      
    0.14                & 0.14              &    0.13           & \textbf{0.15}     & \textbf{0.15}   &
    0.17                & 0.17              &    0.17           & --    &  \textbf{0.18}
    \\
    \textbf{sign-epinions}   & 
    $2.6e4$ & $1.0e5$ &
    0.10                &    --             &   \textbf{0.18}   & \textbf{0.18}     &
    0.40                & 0.34              &    --             & --     &  \textbf{0.47}           &      
    0.50                & 0.51              &    --             & --     & \textbf{0.54}  
    \\
    \textbf{twitter-follows} & 
    $4.0e5$ & $8.4e5$ &
    0.06                & --                &   \textbf{0.07}   & \textbf{0.07}     &     
    0.24                & 0.23              &    --             &   --              & \textbf{0.37} &
    0.53                & \textbf{0.54}     &    --             &   --              & 0.53
    \\
    \bottomrule
    \end{tabular}
    }
    \caption{Coverage rate achieved in \kdDSP at $k = 64$ on real graphs. \texttt{FC} and \texttt{DQN} are short for \model and \texttt{S2V-DQN} respectively. 
    -- denotes failure to terminate within the time limit $t = 900s$.
    }
    \label{tab:real_k_64}
\end{table*}

\begin{figure*}[!ht]
    \centering
    \includegraphics[width=.82\textwidth]{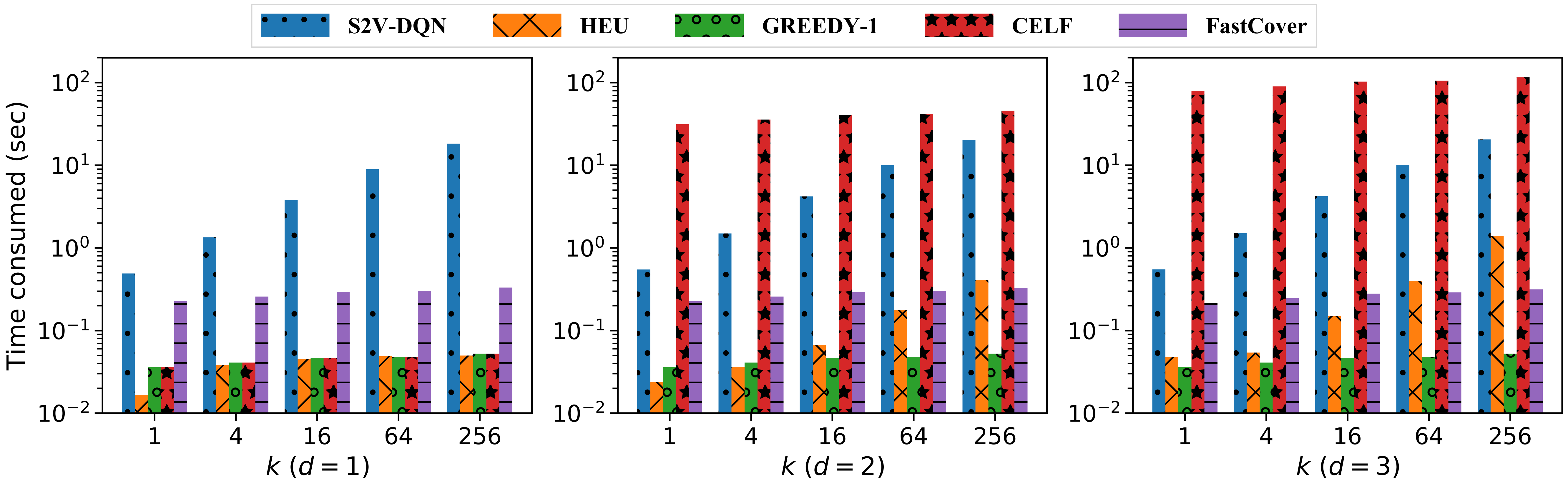}
    \caption{Running time consumed for \kdDSP on \texttt{soc-anybeat} with different $d$}
    \label{fig:running_time}
\end{figure*}

Now we investigate the efficiency of different algorithms. As an illustrating example, we trace the running time consumed by different methods on the graph \texttt{soc-anybeat} in Fig. \ref{fig:running_time}, which has a modest size such that all the algorithms terminate within the time limit.

We distinguish the problem \kdDSP from $d = 1$ from $d = 2, 3$ as they have difference theoretical complexity. When $d = 1$, \texttt{GREEDY-1} (same as \texttt{CELF}) and \texttt{HEU} are most efficient, around 2 to 3 times faster than \model. On the other hand, \texttt{S2V-DQN} has the highest complexity, with running time increasing in $k$ significantly, as the importance of each rode re-evaluated through GNN in each iteration, while \model only makes one forward pass. 

For $d = 2, 3$,  we first empirically verify that the running time of \model, \texttt{DQN} and \texttt{GREEDY-1} is hardly influenced by $d$, which matches the theoretical results in Table \ref{tab:complexity}. On the other hand, \texttt{CELF} and \texttt{HEU} have a much higher time complexity when $d$ increases due to the evaluation of the $d$-neighborhood.
We find that the running time of \model is comparable or even less than \texttt{HEU} when $k$ increases and can be more than $900$ times faster than \texttt{CELF} on \texttt{soc-anybeat}. Similar results on the running time are observed on the other graphs, which demonstrate the efficiency of our method \model.


\section{Conclusion}
\label{sect:conclusion}

In this paper, we focus on influence maximization, an important problem aiming at selecting influential users in social networks to promote propagation with limited budget. We first model the influence propagation process in a deterministic multi-hop approach as a $d$-hop coverage problem. 
Then we propose a novel unsupervised framework \model incorporating graph neural networks to solve the budget-constrained $d$-hop dominating set problem (\kdDSP), which determines the seed set with only one forward pass of GNN, trained with a well-motivated probability-based loss function. We also propose a novel graph network architecture GRAT, outperforming the mainstream GNNs in the specific optimization in \model. The experimental results on synthetic and large real graphs prove that \model computes a near-optimal coverage while achieving significant speedup compared to the benchmark non-ML or ML methods such as \texttt{CELF} and \texttt{S2V-DQN} in practice.

\bibliographystyle{siamplain}
\bibliography{ref}

\end{document}